# Gamma and Vega Hedging Using Deep Distributional Reinforcement Learning[*]

Jay Cao, Jacky Chen, Soroush Farghadani, John Hull, Zissis Poulos, Zeyu Wang, Jun Yuan

December 2022

First Draft: April 2022

## Abstract

We show how D4PG can be used in conjunction with quantile regression to develop a hedging strategy for a trader responsible for derivatives that arrive stochastically and depend on a single underlying asset. We assume that the trader makes the portfolio delta neutral at the end of each day by taking a position in the underlying asset. We focus on how trades in the options can be used to manage gamma and vega. The option trades are subject to transaction costs. We consider three different objective functions. We reach conclusions on how the optimal hedging strategy depends on the trader's objective function, the level of transaction costs, and the maturity of the options used for hedging. We also investigate the robustness of the hedging strategy to the process assumed for the underlying asset.

[*]This work was supported by the Financial Innovation Hub (FinHub), a research center at the Joseph L. Rotman School of Management, University of Toronto. FinHub is funded by the TD Bank, the Global Risk Institute in Financial Services, the Rotman Catalyst Fund, the Royal Bank of Canada, the Bank of Canada, and the University of Toronto's Data Science Institute. All authors are affiliated with FinHub: jay.cao@rotman.utoronto.ca, jackyjc.chen@rotman.utoronto.ca, s.farghadani@mail.utoronto.ca, hull@rotman.utoronto.ca, zissis.poulos@rotman.utoronto.ca, colinzeyu.wang@utoronto.ca, jun.yuan@rotman.utoronto.ca. We are grateful to Mukesh Kayadapuram and Xuanze Li for excellent research assistance.

# Gamma and Vega Hedging Using
# Deep Distributional Reinforcement Learning

## 1. Introduction

Hedging a portfolio of options or more complex derivative securities involves sequential decision making. The position that is taken to reduce risk must be revised daily, or even more frequently, to reflect movements in market variables. Reinforcement learning is therefore an ideal tool for finding a hedging strategy that optimizes a particular objective function.

Prior research concerned with the application of reinforcement learning to hedging decisions includes Halperin (2017), Buehler et al (2019), Kolm and Ritter (2019), and Cao et al (2021). These authors consider how reinforcement learning can be used to hedge a single call or put option using a position in the underlying asset. The key measure of risk they consider is the delta of the option. This is the first partial derivative of the option price with respect to the price of the underlying asset and is therefore a measure of exposure to small changes in the price of the underlying asset. The position taken in the underlying asset can reduce or eliminate this exposure.

In practice, a trader responsible for derivatives dependent on a particular underlying asset does not have a great deal of discretion about delta hedging: the trader is usually required to eliminate, or almost eliminate, the delta exposure each day.[1] A more interesting decision that the trader has to make concerns the gamma and vega exposure. Gamma is the second partial derivative of the value of the portfolio with respect to the underlying asset and is a measure of exposure to large asset price changes. Vega is the partial derivative of the value of the portfolio with respect to the volatility of the asset price and is measure of exposure to volatility changes. A trader is typically subject to limits on the permissible size of a portfolio's gamma and vega exposure, but has discretion on how gamma and vega are managed within those limits. Whereas delta can be changed by taking a position in the underlying asset, gamma and vega can be changed only by taking a position in an option or other derivative.

---

[1] Trading rooms are generally organized so that responsibility for all derivatives dependent on a particular underlying asset is assigned to a single trader.

In deciding on a hedging strategy, traders must take transaction costs into account. These arise because of the difference between the bid price (price at which a financial instrument can be sold to a market maker) and ask price (price at which the financial instrument can be bought from the market maker). The mid-market price (i.e., the average of the bid and ask prices) is usually considered to be the true value of the instrument. As a result, the transaction cost can be quantified as half the difference between the bid and ask prices. Transaction costs are typically quite small for trading the underlying asset, but much larger for trading options and other derivatives.

Rather than considering a single option, we consider a portfolio of options that evolves stochastically. We assume that the portfolio is made delta-neutral at the end of each day. Before a final trade in the underlying asset is made to achieve delta neutrality, the trader makes a decision on how gamma and vega are managed. As a simplification, we assume that there are no transaction costs associated with trading the underlying asset. (As mentioned earlier these costs are usually quite small.) Alternative levels for transaction costs for the options brought into the portfolio for hedging are considered.

Our paper can be regarded as a "proof of concept". We make relatively simple assumptions about the process driving the arrival of client options, the options available for hedging, the process driving the price of the underlying asset, the nature of transaction costs, and the trader's objective function. Client options are assumed to arrive according to a Poisson process with intensity 1.0 per day. The options entered into by clients are assumed to have 60 days to maturity and be equally likely to be long or short. The option available for hedging each day are assumed to be at-the-money call with a particular time to maturity.[2] Each day the trader chooses the position (if any) to be taken in this option. The trader then trades the underlying asset to make the whole portfolio (options entered into with clients plus options entered into for hedging) delta-neutral. The transaction costs associated with an option trade are assumed to be a specified proportion of the value of that trade. The performance of the hedging strategy is considered over 30 days. At the end of that period, all options in the portfolio are valued at their mid-market prices.

---

[2] At-the-money options are efficient hedging instruments because they have a large gamma and vega compared with similar maturity options that are significantly in- or out-of-the-money.

We consider three different objective functions. The first involves a trade-off between the mean and standard deviation of the trader's profit or loss. The second involves minimizing value at risk (i.e., minimizing a particular percentile of the loss distribution). The third involves minimizing conditional value at risk (also referred to as expected shortfall), which is the expected loss conditional that it is worse than a particular percentile of the loss distribution.

Our approach can be used to assess the profitability of trading derivatives when a particular hedging strategy is used. The expected profit on client options (i.e., the expected transaction costs earned on these options) can be compared with the expected loss on hedging (i.e., the expected transaction costs paid on the options used for hedging). Assuming the former are calculated in the same way as the latter, we find that the hedging strategies for the cases we consider are economically feasible in the sense that expected transaction costs earned are greater than those paid.

The rest of the paper is organized as follows. Section 2 describes the reinforcement learning model we use. Section 3 explains the models we use for the evolution of the underlying asset price. Section 4 presents our illustrative results. It first focuses on gamma hedging by using a model where volatility is constant and then considers a stochastic volatility world where both gamma and vega have to be monitored. Section 5 investigates the robustness of the reinforcement learning strategies to the assumptions made about the asset price process. Conclusions and suggestions for further work are in Section 6.

## 2. The RL Model

In the standard RL formulation, the goal of the agent is to maximize expected future rewards. Specifically, the agent attempts to maximize at time *t* the expected value of $G_t$, where

$$G_t = R_{t+1} + \gamma R_{t+2} + \gamma^2 R_{t+3} + \cdots + \gamma^{T-1} R_T$$

Here $R_t$ is the reward at time t, $T$ is a horizon date, and $\gamma \in (0,1]$ is a discount factor. Define $Q(S,A)$ as the expected value of $G_t$ from taking action $A$ in state $S$ at time $t$. The standard RL formulation involves maximizing $Q(S,A)$. This is not appropriate for our application (and many other applications in finance) because we are concerned with exploring risk-return trade-offs. A class of RL methods collectively referred to as risk-aware reinforcement learning has recently been developed to extend reinforcement learning so that other attributes of the distribution of $G_t$ beside the mean can be considered.

Tamar et al (2016) show how both the first and second moment (and possibly higher moments) of the distribution of $G_t$ can be updated using more than one Q-function. Cao et al (2021) used this approach and produced results for an objective function involving the mean and standard deviation of $G_t$, where one Q-function approximates the mean of the terminal return and a second Q-function approximates the expected value of the square of the terminal return. However, the approach is less than ideal. It is computationally demanding and imposes restrictions on the objective functions that can be considered.

Bellemare et al (2017) have pointed out that the Bellman equation used to update $Q(S,A)$ can also be used to update distributions. Define $Z(S,A)$ as the distribution of $G_t$ resulting from action $A$ in state $S$. They suggest a procedure known as C51 where $Z(S,A)$ is defined as a categorical distribution and where probabilities are associated with 51 fixed values of $G_t$. As new points on the distribution are determined in trials they are allocated to the neighboring fixed points. Barth-Maron et al (2018) incorporate this approximation method into an actor-critic RL model, where one neural network (critic) estimates the categorical distribution $Z(S,A)$ and another neural network (actor) decides the actions that the agent takes based on information from the critic network. These two neural networks are trained simultaneously. Their framework also supports the use of multiple agents at the same time for distributed exploration of the search space in large-scale RL problems. The overall process is known as Distributed Distributional Deep Deterministic Policy Gradients (D4PG) and has become an impactful RL algorithm with applications primarily in robotic control systems.

Dabney et al (2018) proposed an alternative distributional RL algorithm, QR-DQN, that uses quantile regression (QR) to approximate $Z(S, A)$. The distribution is represented with a discrete set of quantiles whose positions are adjusted during training. Compared to C51, QR generalizes better and is more flexible. It can efficiently approximate a wide range of distributions without the need to specify fixed points a priori. The QR procedure is wrapped within Deep Q-Learning (DQN), a well-known and studied RL algorithm. Unlike actor-critic architectures, DQN uses only a single neural network. The neural network approximates $Z(S, A)$ and the agent actions are generated using a greedy algorithm. The authors show that QR produces better results than C51 when both used in DQN algorithms and applied to Atari 2600 games.

For the hedging problem at hand QR is an attractive solution as it is straightforward to measure value at risk (VaR) and conditional value at risk (CVaR) directly on the quantile-based representation that it generates. In contrast, C51 requires interpolating on the categorical distribution to obtain quantiles, which induces additional approximation errors. Further, an actor-critic architecture similar to the one used in D4PG is also more desirable compared to DQN. The latter is a lightweight sample-efficient model, often used in cases where simulation is slow. In our setting, simulating the portfolio is a fast process and not the bottleneck, so that actor-critic models can handle the task well while exploring the search space rigorously. We posit that this is an important step toward identifying hedging strategies that are sensitive to different hedging scenarios and volatility movements in the underlying.

The combination of QR and D4PG has not been explored in the literature. In this work, as one of our contributions, we modify D4PG to support QR at the output of the critic neural network. The resulting algorithm, which we refer to as D4PG-QR, is used for all experiments discussed later. We observed that our implementation is superior to D4PG (which uses C51 by default) for VaR and CVaR objective functions and as good as D4PG for the mean/standard deviation objective in terms of accuracy and computational efficiency.[3]

---

[3] We use D4PG with a single agent in all experiments since the distributed version is not necessary.

We now describe how we use the distributional RL framework for hedging. We assume that we are hedging a portfolio of client options. The portfolio composition evolves as new client options arrive. Arrivals are modeled with a Poisson process the intensity of which can be specified as a parameter in our framework. In expectation, half of the client orders are short and half are long. At the time of each rebalancing, we use an at-the-money call option for hedging. We assume that the trader rebalances at time intervals of $\Delta t$.

Prices for the underlying asset are generated by a pre-specified stochastic process, which will be described in the next section. All options are assumed to give the holder the right to buy or sell 100 units of the underlying asset. We set $\Delta t$ equal to one day and the time period, $T$, considered is 30 days. We set $\gamma = 1$ as this period is fairly short.

The state at time $i\Delta t$ is defined by:
- The price of the underlying asset.
- The gamma of the portfolio.
- The vega of the portfolio.
- The gamma of the at-the-money option used for hedging.
- The vega of the at-the-money option used for hedging.

The portfolio gamma is calculated as the sum of the gammas of all the options in the portfolio. Portfolio vega is calculated similarly.

The action at time $i\Delta t$ is the proportion of maximum hedging that is done. Specifically, we do not allow the agent to try any arbitrary position in the hedging option during training. Instead, we determine at each time step the maximum hedge permitted such that at least one of the two following criteria is satisfied: (a) the ratio of gamma after hedging to gamma before hedging falls in the range [0,1] and (b) the ratio of vega after hedging to vega before hedging falls in the range [0,1]. The action of the agent is constrained to lie within the resulting range.[4] This ensures that the agent is hedging rather than speculating. Restricting the action space increases sample efficiency and improves convergence. We avoid wasteful simulations where the agent tries actions that are known not to be part of any optimal hedging strategy and therefore reduce the size of the search space.

---

[4] This can be done by choosing an appropriate activation function in the last layer of the neural network that models the agent's action.

Define:

$V_i$:    The value of the option used for hedging at time $i\Delta t$.

$H_i$:    The position taken in the option used for hedging at time $i\Delta t$.

$\kappa$:    The transaction cost associated with the option used for hedging as a proportion of the value of the option.

$P_i$:    The total value of options in the portfolio at time $i\Delta t$ that have not previously expired.

The variable $P_i$ includes all the options in the portfolio that have not expired before time $i\Delta t$. If an option expires at time $i\Delta t$, its value at time $i\Delta t$ is set equal to its intrinsic value.

The reward at time $i\Delta t$ ($i > 0$) is therefore[5]

$$R_i = -\kappa|V_i H_i| + (P_i - P_{i-1})$$

## 2.1 Hedging with D4PG-QR

The RL framework is provided in Figure 1 and it includes the proposed D4PG-QR architecture. There are three major components: the environment, an actor neural network (NN) and a critic NN. The trading environment involves a simulator of how the portfolio composition and value evolves over time based on the hedging positions that the agent takes, the market dynamics for the underlying, and the client option arrival process. At any given time, the simulator computes the next state and the reward. The actor NN (also known as policy network) implements the hedging strategy. It takes as input a state and outputs the position of the hedger. The critic NN takes as inputs a state, $S$, and the action from the actor's output, $A$. Its role is to (a) estimate the distribution of the trading loss at the end of the hedging period, $Z(S, A)$, when taking action $A$ in state $S$, and (b) compute gradients that minimize the objective function $f(Z(S, A))$. These gradients are then used by the actor NN to improve the actions that it outputs in subsequent rounds.

---

[5] As mentioned, we assume that delta hedging can be carried out costlessly. Our $\gamma = 1$ assumption is equivalent to assuming no funding costs and is not unreasonable given the short (30 day) time horizon considered.

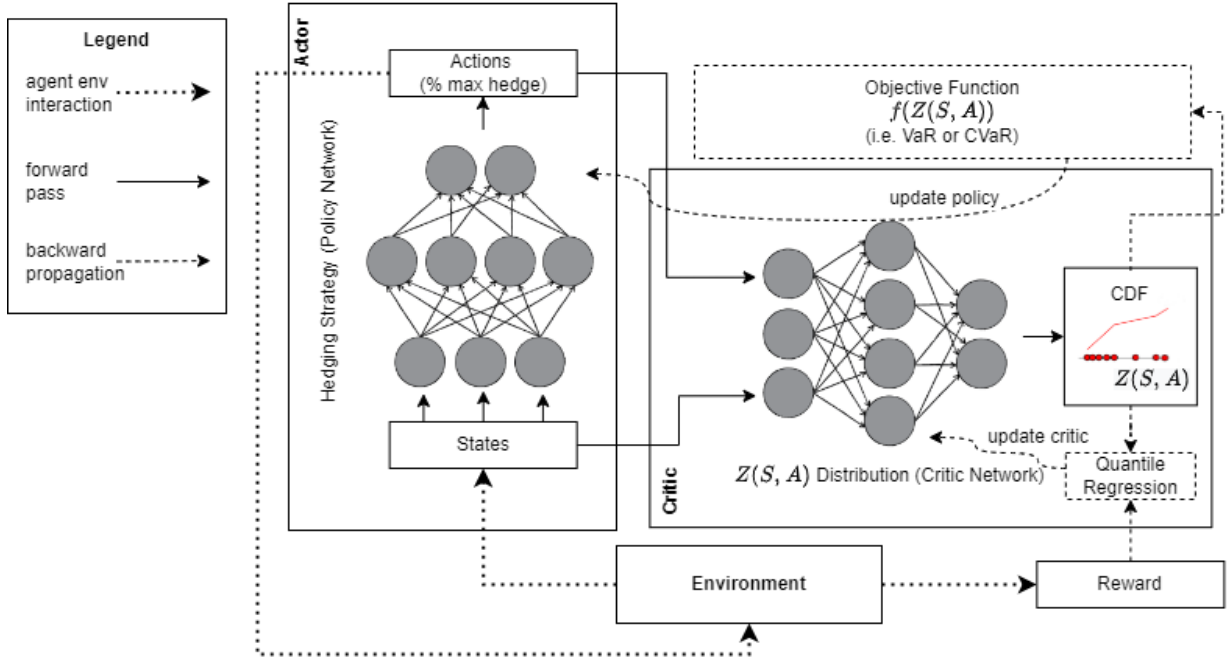

**Figure 1**: RL architecture for actor-critic learning and proposed networks.

The key difference between D4PG and our architecture is that $Z(S, A)$ is represented by a set of discrete quantiles, whose probabilities are fixed but their position is adjusted during training based on the rewards observed over several episodes. This quantile-based representation also allows us to compute the CDF of $Z(S, A)$. The adjustment is done via quantile regression using the quantile Huber loss as a loss function. The Huber loss function is defined as:

$$L_k(u) = \begin{cases} \frac{1}{2}u^2, & |u| \leq k \\ k\left(|u| - \frac{1}{2}k\right), & otherwise \end{cases}$$

The quantile Huber loss used as the critic's loss in D4PG-QR corresponds to an asymmetric squared loss in the interval $[-k, k]$ and acts as a regular quantile loss outside that interval. Specifically, the critic's quantile Huber loss for any quantile $\tau \in [0,1]$ and parameter $k$ is:

$$h_\tau^k(u) = |\tau - \delta_{u<0}| L_k(u)$$

where $\delta_{u<0}$ denotes a Dirac at all negative points.

Gradients that minimize the loss are used to improve the critic's accuracy when estimating $Z(S, A)$. In early training stages the critic's estimate of $Z(S, A)$ is very rough and thus the actor's hedging policy is far from optimal. As the critic improves so does the actor's policy to the point where both NNs converge to some local optimum: the critic ends up with a good estimate of the distribution of hedging loss for a large range of states and actions and the actor learns a good policy that optimizes the corresponding objective for any state that it is presented with.

We should point out that there are several other components (not shown in Figure 1) that we borrow from the D4PG algorithm, the most important being the use of a replay buffer. With a replay buffer, instead of learning a policy using state transitions in the order that they appear based on the actor's actions, we simulate and store a large set of state transitions (along with their rewards) in a table. Then, to update the critic and actor NNs, we sample transitions and their rewards uniformly at random from that table. This aids with convergence of the NNs for two reasons. First, the samples are i.i.d. which is beneficial when updating NNs in general. Second, the replay buffer allows for "experience replay". That is, we can use the same state transition and reward multiple times when updating critic and actor. Since the improvements on estimating $Z(S, A)$ are incremental, applying gradient updates several times over the same state transitions leads to faster convergence. The pseudocode (Algorithm 1) describing the parameter update process for the actor and critic networks in D4PG-QR is given below:

---

**Algorithm 1** D4PG-QR

**Input:** batch size $N$, quantile representation $Z_w^{\tau_j}$ of $Z_w$ for $j = 1...M$, initial learning rates $\alpha_0$ (actor, $\pi_\theta$) and $\beta_0$ (critic, $Z_w$)

1: Initialize actor parameters $\theta$ and critic parameters $w$ at random
2: **for** $t = 1 : T$ **do**
3:     Sample $N$ transitions $(S_i, A_i, R_i) \rightarrow (S_{i+1}, A_{i+1}, R_{i+1})$ from replay buffer
4:     Compute target distribution $Y_i \leftarrow R_i + \gamma Z_w(S_{i+1}, \pi_\theta(S_{i+1}))$
5:     Compute the expected quantile Huber loss $L_i(w) \leftarrow \sum\limits_{j=1}^{M} \mathbb{E}_{Y_i}[h_{\tau_j}^k(Y_i - Z_w^{\tau_j}(S_i, A_i))]$
6:     Compute actor and critic parameter updates:
7:         $\Delta_w \leftarrow \frac{1}{N} \sum\limits_{i=1}^{N} \nabla_w L_i(w)$
8:         $\Delta_\theta \leftarrow \frac{1}{N} \sum\limits_{i=1}^{N} \nabla_\theta \pi_\theta(S_i) \mathbb{E}[\nabla_A f(Z_w(S_i, A))]|_{A=\pi_\theta(S_i)}$
9:     $w \leftarrow w + \beta_t \Delta_w$
10:     $\theta \leftarrow \theta + \alpha_t \Delta_\theta$

## 3. The Asset Pricing Model

We assume that the risk-neutral behavior of the underlying asset price, $S$, and its volatility, $\sigma$, are governed by the following stochastic processes.[6]

$$dS = (r - q)S\,dt + \sigma S\,dz_1$$

$$d\sigma = v\sigma\,dz_2$$

where $dz_1$ and $dz_2$ are Wiener processes with constant correlation $\rho$. The volatility of volatility variable, $v$, the risk-free rate, $r$, and the dividend yield, $q$, are assumed constant. We assume that the real-world expected return $\mu$ is also constant. The real-world process for the asset price is the same as that given above with $r - q$ replaced by $\mu$.

This model is a particular case of the SABR model developed by Hagen et al (2002).[7] It has the attractive property that there is a good analytic approximation to a European option's implied volatility. Define $\sigma_0$ and $S_0$ as the initial values of $\sigma$ and $S$. If the strike price and time to maturity of a European call option are $K$ and $T$, the estimate of the implied volatility is $\sigma_0 B$ when $F_0 = K$ and $\frac{\sigma_0 B \phi}{\chi}$ otherwise where

$$F_0 = S_0 e^{(r-q)T}$$

$$B = 1 + \left( \frac{\rho v \sigma_0}{4} + \frac{(2 - 3\rho^2)v^2}{24} \right) T$$

$$\phi = \frac{v}{\sigma_0} \ln\left(\frac{F_0}{K}\right)$$

$$\chi = \ln\left( \frac{\sqrt{1 - 2\rho\phi + \phi^2} + \phi - \rho}{1 - \rho} \right)$$

---

[6] For the rest of the article, $S$ refers to the asset price rather than the state.
[7] The general SABR model is (in a risk-neutral world) $dF = \sigma F^\beta dz_1$ with $d\sigma = v\sigma dz_2$ where $F$ is the forward price of the asset for a particular maturity. We set $\beta = 1$ and assume that the volatility, $\sigma$, applies to the evolution of all forward prices. When the forward contract matures at time $T$, $S = Fe^{-(r-q)(T-t)}$ so that $S$ follows the process indicated.

Denoting the implied volatility by $\sigma_{imp}$, the value of the option is given by the Black-Scholes-Merton formula:

$$S_0 N(d_1) e^{-qT} - K e^{-rT} N(d_2) \quad (2)$$

where

$$d_1 = \frac{\ln\left(\frac{S_0}{K}\right) + \left(r - q + \frac{\sigma_{imp}^2}{2}\right) T}{\sigma_{imp} \sqrt{T}}$$

$$d_2 = d_1 - \sigma_{imp} \sqrt{T}$$

and $N$ is the cumulative normal distribution function. When $v = 0$ the implied volatility is constant and equal to $\sigma_0$ and the SABR model reduces to the option pricing model developed by Black and Scholes (1973) and Merton (1973).

As mentioned, delta is the first partial derivative of the option price with respect to the asset price, gamma is the second partial derivative with respect to the asset price, and vega is the first partial derivative with respect to volatility. Under the model that is being assumed, a natural idea is to regard a European option value as a function of $S$ and $\sigma$ for the calculation of these partial derivatives. However, the usual practitioner approach is to regard the option value as a function $S$ and $\sigma_{imp}$ when Greek letters are calculated. This is the approach we will adopt. Denoting delta, gamma, and vega by $\Delta$, $\Gamma$, and $Y$ respectively, equation (2) gives:

$$\Delta = N(d_1) e^{-qT}$$

$$\Gamma = \frac{N'(d_1) e^{-qT}}{S_0 \sigma_{imp} \sqrt{T}}$$

$$Y = \sigma_{imp} \sqrt{T} N'(d_1)$$

The delta, gamma, and vega of a portfolio are calculated by summing those for the individual options in the portfolio.

## 4. Results

We conducted the hedging experiment with three different objective functions. The objective functions are calculated from the total gain/loss during the (30-day) period considered. The first objective function to be minimized is

$$m + 1.645s$$

where $m$ is the mean of the loss and $s$ is its standard deviation.[8] If the distribution of losses were normal, this objective function would minimize the 95$^{th}$ percentile of the portfolio loss distribution.

The second and third objective functions minimize the value at risk and conditional value at risk. For both measures we use a 95% confidence level. Our value at risk measure (VaR95) is therefore the 95$^{th}$ percentile of the portfolio loss while our conditional value at risk measure (CVaR95) is the expected loss conditional on the loss being worse than the 95$^{th}$ percentile of the loss distribution.

The results we report are averages over 5,000 test scenarios for a 30-day hedging period. The test scenarios are different from the (much greater number of) scenarios on which agents are trained. Within each set of tests, the scenarios were kept the same.

**4.1 Hedge Ratios**

The gamma (vega) hedge ratio when a hedging decision is made is the proportional amount by which gamma (vega) is reduced. It is defined as one minus the gamma (vega) exposure of the portfolio after the hedging divided by the gamma (vega) exposure of the portfolio before hedging, i.e.:

$$\text{Gamma Hedge Ratio} = 1 - \frac{\sum_t sign(\Gamma^t)\Gamma^{t^+}}{\sum_t sign(\Gamma^t)\Gamma^t}$$

$$\text{Vega Hedge Ratio} = 1 - \frac{\sum_t sign(\Upsilon^t)\Upsilon^{t^+}}{\sum_t sign(\Upsilon^t)\Upsilon^t}$$

---

[8] Gains are regarded as negative losses. When gamma and vega are hedged, $m$ is positive because we do not take account of the profit from the options being hedged.

where $\Gamma^t$ is portfolio gamma before hedging at time t and $\Gamma^{t^+}$ is this ratio after hedging. Similarly, $\Upsilon^t$ and $\Upsilon^{t^+}$ are the portfolio vega before and after hedging. The values we report are the average values of the gamma and vega hedge ratios across all test scenarios and all hedging actions.

## 4.2 Gamma Hedging Results

We start by focusing on gamma risk by setting $v$ = 0 so that the SABR model reduces to the Black-Scholes-Merton model and there is no vega risk. We remove the two vega-related state variables in this experiment as they are not relevant to the agent. As already mentioned, we assume the RL agent has to hedge the client orders arriving according to Poisson process with intensity 1.0 per day. Each client order is assumed to be for a 60-day option on 100 units of the underlying assets and has an equal probability of being long or short. We run experiments with transaction cost assumed to be 0.5%, 1% and 2% of the option price. A 30-day at-the-money option is used as the hedging option. The initial stock price is set at $10 and the volatility is 30% per annum.[9]

Table 1 compares the performance of RL agents with delta-neutral and delta-gamma-neutral strategies when they use the three different objective functions and the three different transaction cost assumptions. It shows that RL improves the hedger's objective functions compared to the simpler strategies.[10] The outperformance of the RL agents can be attributed to their ability to adjust their hedging policies for different transaction costs. As the transaction cost increases, the RL agents reduce the amount of gamma they are hedging. The expected cost of hedging increases as the transaction cost increases. However, because less hedging is done the expected cost of hedging rises more slowly than transaction costs.

The value of one client option is about $60 based on Black-Scholes pricing formula with our chosen parameters. With the expected arrival of one option per day, the expected profit from client options over 30 days is therefore about $1,800 times the premium over the mid market value charged on client options. If this premium is $\kappa$ times the option price (i.e. the transaction cost faced by clients of the dealer is the same as the transaction cost faced by the dealer when hedging), the results in the table show that the cost of hedging is comfortably covered by the transactions costs earned on client options. For example, when $\kappa$ $is$ 1% the expected cost of the RL hedge is less than $5 whereas the expected profit from client options is $18.

---

[9] The results presented are for $r = q = \mu = 0$. We tried a range of other values for these parameters and results were similar to those presented.
[10] We obtain good results in almost all our experiments, but there is no theory guaranteeing convergence to the optimal strategy.

**Table 1:** Results of tests when volatility is constant. The Delta column shows the values of the objective functions when only delta hedging is carried out; the Delta-Gamma column shows the values when delta and gamma are fully hedged. The RL results column shows the values when RL agents are used to minimize the objective functions in the first column. The final two columns report the average percentage of gamma hedged by the RL agent and the expected loss arising from transaction costs from the trading conducted by the RL agent.

| Objective Function | Objective Function Value for | | | RL Gamma Hedge Ratio | Expected RL Transaction Cost |
|---|---|---|---|---|---|
| | Delta | Delta-Gamma | RL | | |
| **0.5% Transaction Cost** | | | | | |
| Mean-Std | 24.61 | 5.78 | 5.44 | 0.83 | 3.00 |
| VaR95 | 24.29 | 5.78 | 5.47 | 0.75 | 2.80 |
| CVaR95 | 36.64 | 7.13 | 6.78 | 0.79 | 2.92 |
| **1% Transaction Cost** | | | | | |
| Mean-Std | 24.61 | 9.93 | 8.36 | 0.57 | 4.58 |
| VaR95 | 24.29 | 10.12 | 8.63 | 0.56 | 4.51 |
| CVaR95 | 36.64 | 11.55 | 10.02 | 0.60 | 4.71 |
| **2% Transaction Cost** | | | | | |
| Mean-Std | 24.61 | 18.74 | 12.73 | 0.30 | 5.91 |
| VaR95 | 24.29 | 19.11 | 13.05 | 0.24 | 5.15 |
| CVaR95 | 36.64 | 21.10 | 15.37 | 0.29 | 5.78 |

Figure 2 shows that the distribution of the gain for the delta, delta-gamma, and RL strategies for the VaR95 agent when transaction costs are 1%. An examination of the left tails of the distributions in the rug plot shows that the VaR95 RL strategy has a smaller probability of experiencing large losses than the delta and delta-gamma strategies. The expected cost of delta hedging is zero. (This is as expected because we assume no transaction costs for delta hedging.) The expected cost of RL is clearly less than delta-gamma strategy.

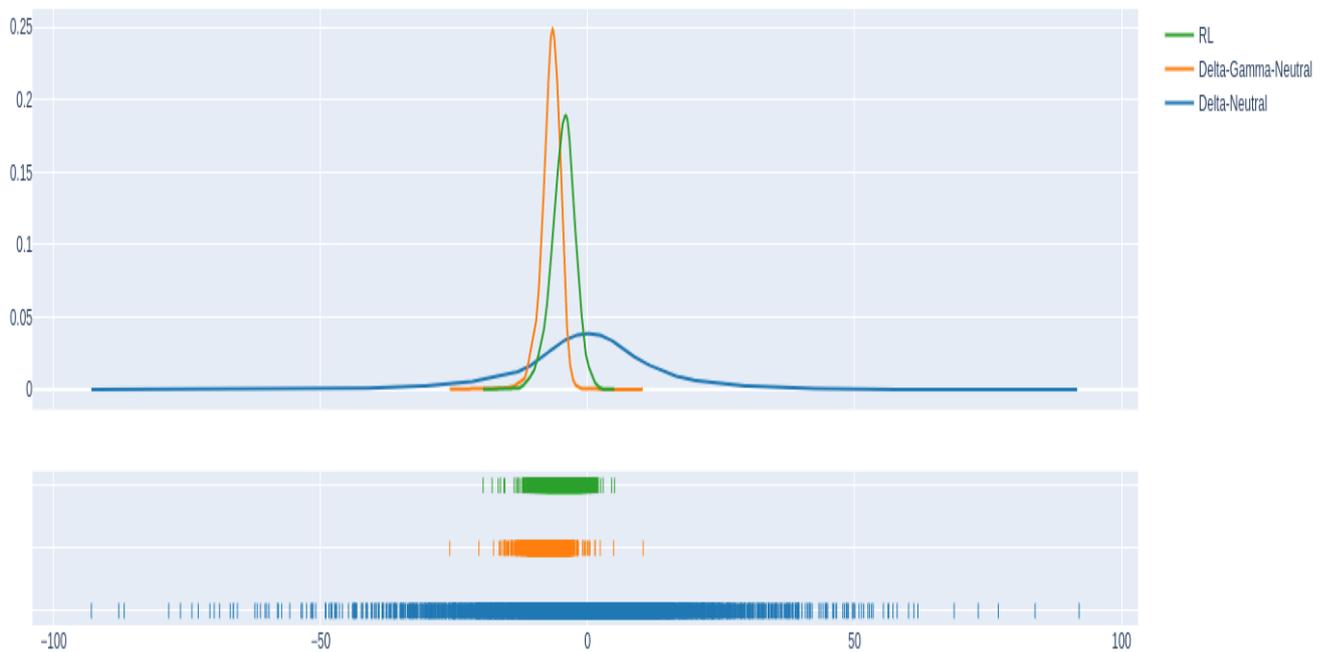

**Figure 2:** Comparison of gain distribution for delta hedging, delta-gamma-neutral hedging, and hedging by the VaR95 RL agent when transaction cost = 1%. Note that the 5$^{th}$ percentile of the gain distribution below corresponds to the 95$^{th}$ percentile loss in Table 1

**4.3 Risk Limits**

Traders are usually subject to limits on their gamma and vega exposure at the end of each day. These limits can easily be incorporated into RL by imposing constraints, in addition to those mentioned in Section 2, on the range of hedging transactions that are considered considered.

As an experiment, we compared RL with delta-gamma hedging when gamma is reduced only when the exposure exceeds a risk limit. We choose a risk limit equal to 10% of the max gamma exposure of the unhedged portfolio in the simulated environment, which could be an appropriate choice for a trading desk to set the risk limit in practice. As Table 2 shown, the RL agent continues to outperform the rule-based agents across different objective functions and transaction costs.

**Table 2:** Results of tests when volatility is constant. The Delta-Gamma agent shows the values when delta is fully hedged, and gamma is fully hedged every time the dollar gamma exposure is greater than 50. The RL results show the values when delta is fully hedged and gamma exposure is hedged according to the VaR95 RL agent policy when the dollar gamma exposure is greater than 50.

| Objective Function | Objective Function Value for | |
|---|---|---|
| | Delta-Gamma | RL |
| **0.5% Transaction Cost** | | |
| Mean-Std | 7.27 | 6.80 |
| VaR95 | 7.40 | 6.43 |
| CVaR95 | 9.20 | 7.84 |
| **1.0% Transaction Cost** | | |
| Mean-Std | 9.59 | 9.19 |
| VaR95 | 9.69 | 8.86 |
| CVaR95 | 11.62 | 10.39 |
| **2.0% Transaction Cost** | | |
| Mean-Std | 14.85 | 13.34 |
| VaR95 | 15.17 | 13.29 |
| CVaR95 | 17.44 | 15.33 |

**4.4 Impact of Volatility Uncertainty**

We now move on to consider situations where volatility is uncertain by setting $\sigma_0$ equal to 30%, $v$ equal to 0.3, and $\rho = -0.7$ in the SABR model described in Section 3. Other parameters and assumptions are as in Section 4.2. With two options, both gamma and vega could be completely neutralized. However, we assume that only a single at-the money option is available. Whereas short maturity at-the-money options are most useful for hedging gamma, longer maturity options work better for vega. Table 3 illustrates this by showing results for two different maturities of the options used for hedging: 30 days and 90 days. As before three different transaction costs, as a percent of the hedging option price, are considered. The hedging policy of the RL agents is adaptive to the maturity of the hedging options. The performance of the RL agents is closest to the delta-gamma-neutral policy when short-dated maturity options are used for hedging and the performance moves closer to the delta-vega neutral policy as the option maturity increases. Similar to the results when the Black-Scholes-Merton model is used, the RL agents reduce the amount of gamma and vega hedging as transaction cost increases. As for Table 1, it can be shown that the expected cost of hedging paid is comfortably covered by the transaction costs earned on client options in the situations we consider.

**Table 3:** Results of tests when volatility is stochastic. The Delta-column shows the values of the objective function when only delta hedging is carried out; the Delta-Gamma column shows the values when delta and gamma are fully hedged; the Delta-Vega column shows the values when delta and vega are fully hedged. The RL results column shows the values when RL agents are used to minimize objective functions. The final three columns report the averages of the gamma and vega hedged by RL and the expected RL loss from transaction costs.

| Hedge Option Maturity | Objective Function | Value of Objective Function for | | | | RL Gamma Hedge Ratio | RL Vega Hedge Ratio | Expected RL Trans-action Cost |
|---|---|---|---|---|---|---|---|---|
| | | Delta | Delta-Gamma | Delta-Vega | RL | | | |
| **Transaction Costs = 0.5%** | | | | | | | | |
| 30 days | Mean-Std | 35.76 | 19.46 | 44.82 | 17.76 | 0.53 | 0.17 | 2.75 |
| | VaR95 | 34.43 | 19.23 | 42.90 | 19.31 | 0.61 | 0.18 | 3.25 |
| | CVar95 | 52.91 | 27.53 | 62.77 | 26.06 | 0.73 | 0.16 | 3.44 |
| 90 days | Mean-Std | 35.76 | 25.25 | 15.47 | 14.28 | 0.27 | 0.55 | 4.90 |
| | VaR95 | 34.43 | 24.43 | 15.40 | 14.41 | 0.24 | 0.47 | 4.53 |
| | CVar95 | 52.91 | 31.96 | 20.21 | 18.17 | 0.26 | 0.63 | 5.04 |
| **Transaction Costs = 1%** | | | | | | | | |
| 30 days | Mean-Std | 35.76 | 23.06 | 51.36 | 20.03 | 0.50 | 0.13 | 4.40 |
| | VaR95 | 34.43 | 23.02 | 50.24 | 20.22 | 0.45 | 0.12 | 3.98 |
| | CVar95 | 52.91 | 31.55 | 69.92 | 26.81 | 0.42 | 0.10 | 4.00 |
| 90 days | Mean-Std | 35.76 | 35.29 | 22.20 | 18.61 | 0.14 | 0.37 | 6.21 |
| | VaR95 | 34.43 | 35.01 | 22.05 | 18.86 | 0.15 | 0.32 | 6.36 |
| | CVar95 | 52.91 | 42.63 | 27.18 | 24.58 | 0.12 | 0.31 | 5.65 |
| **Transaction Costs = 2%** | | | | | | | | |
| 30 days | Mean-Std | 35.76 | 30.51 | 64.77 | 24.17 | 0.33 | 0.11 | 7.01 |
| | VaR95 | 34.43 | 30.67 | 64.56 | 23.85 | 0.29 | 0.07 | 5.70 |
| | CVar95 | 52.91 | 39.79 | 84.38 | 31.57 | 0.27 | 0.07 | 5.62 |
| 90 days | Mean-Std | 35.76 | 56.67 | 36.58 | 25.20 | 0.09 | 0.21 | 9.31 |
| | VaR95 | 34.43 | 56.97 | 36.78 | 25.73 | 0.08 | 0.16 | 7.80 |
| | CVar95 | 52.91 | 65.05 | 42.14 | 32.62 | 0.09 | 0.16 | 8.13 |

By training multiple RL agents with different VaR percentiles as objective functions and using a 30 days maturity hedging option, we have constructed a frontier to represent the risk and return for different levels of risk aversion when transaction cost is equal to 1% in Figure 3. The figure shows that the RL agents generally outperform simple rule-based strategies such as delta-neutral, delta-gamma-neutral and delta-vega-neutral.

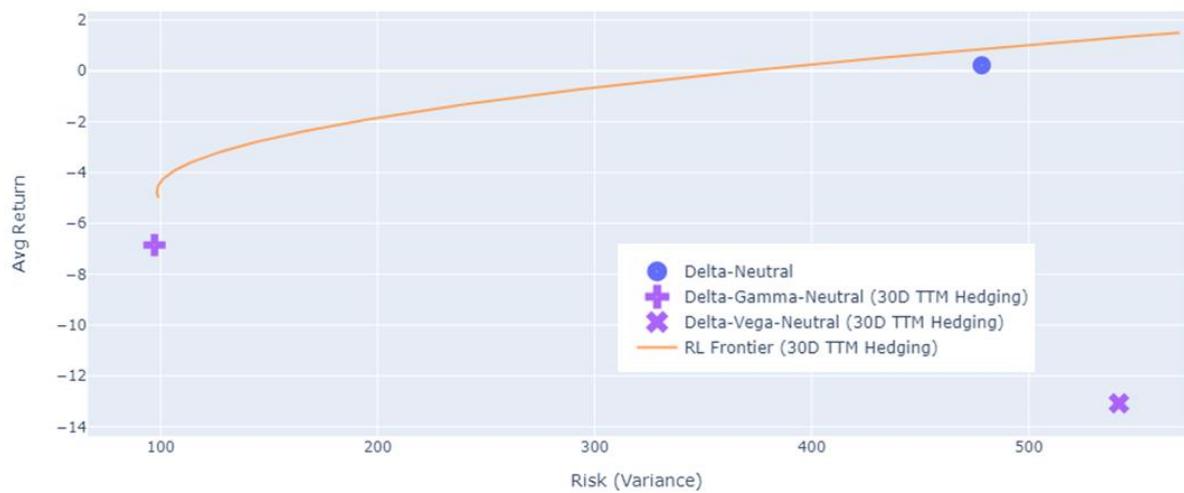

**Figure 3:** Risk-return trade-offs that are possible using RL with different objective functions. The positive average cost of hedging reported in previous tables means the average return of the agent is negative

## 5. Robustness Tests

Our tests so far have assumed that the agent correctly estimates the stochastic process followed by the underlying asset. We now consider the sensitivity of hedge performance to a stochastic process that is different from the one assumed in developing the hedging strategy. We assume that the true stochastic process is the one considered in Table 3 where the volatility of volatility parameter, $v$, is 0.3 and the initial volatility $\sigma$ is 30%. The option used for hedging is assumed to last for 90 days and the transaction costs are 1% of the option price.

**Table 4**: Values of the objective function and expected cost of hedging when the hedging strategy is developed with values of the volatility of volatility, $v$, that are different from the actual 0.3 value. The initial volatility, $\sigma$, is assumed to be 30%.

| $v$ | Objective Function | Value of Objective Function | Gamma Hedge Ratio | Vega Hedge Ratio | Expected Transaction Costs |
|---|---|---|---|---|---|
| 0.0 | Mean-Std | 23.29 | 0.17 | 0.16 | 6.03 |
| | VaR95 | 23.96 | 0.18 | 0.16 | 6.43 |
| | CVaR95 | 32.79 | 0.21 | 0.17 | 7.08 |
| 0.15 | Mean-Std | 21.49 | 0.15 | 0.17 | 6.13 |
| | VaR95 | 20.98 | 0.11 | 0.18 | 4.99 |
| | CVaR95 | 25.86 | 0.17 | 0.24 | 6.66 |
| 0.3 | Mean-Std | 18.61 | 0.14 | 0.37 | 6.21 |
| | VaR95 | 18.86 | 0.15 | 0.32 | 6.36 |
| | CVaR95 | 24.58 | 0.12 | 0.31 | 5.65 |
| 0.45 | Mean-Std | 18.72 | 0.19 | 0.48 | 7.60 |
| | VaR95 | 18.73 | 0.16 | 0.41 | 7.08 |
| | CVaR95 | 22.97 | 0.19 | 0.40 | 7.71 |
| 0.6 | Mean-Std | 19.66 | 0.20 | 0.62 | 9.06 |
| | VaR95 | 18.47 | 0.17 | 0.48 | 7.46 |
| | CVaR95 | 23.48 | 0.20 | 0.52 | 8.74 |

Table 4 considers the situation where the agent develops the hedging strategy with the correct value of $\sigma$ but with values of $v$ equal 0, 0.15, 0.3, 0.45, and 0.6. Table 5 considers the situation where the correct value of $v$ is used but the hedging strategy is developed with values of $\sigma$ equal to 10%, 20%, 30%, 40% and 50%. Overall the results indicate that hedging performance is fairly robust to the values assumed for the parameters. Interestingly, in both cases there is virtually no deterioration in the performance of the hedge when the parameters used to determine the hedging strategy are too high, but a noticeable decrease in performance when they are too low. However, the expected cost of hedging is greater when the parameter estimates are too large.

**Table 5**: Values of the objective function when the hedging strategy is developed with values of the initial volatility, $\sigma$, that are different from the actual 30% value. The volatility of volatility parameter, $v$, is assumed to be 0.3.

| $\sigma$ | Objective Function | Value of Objective Function | Gamma Hedging Ratio | Vega Hedging Ratio | Expected Loss From Trading |
|---|---|---|---|---|---|
| 10% | Mean-Std | 26.57 | 0.28 | 0.44 | 11.12 |
| | VaR95 | 28.58 | 0.29 | 0.34 | 10.40 |
| | CVaR95 | 29.81 | 0.36 | 0.39 | 11.93 |
| 20% | Mean-Std | 21.48 | 0.21 | 0.50 | 8.94 |
| | VaR95 | 19.61 | 0.22 | 0.53 | 8.37 |
| | CVaR95 | 27.07 | 0.18 | 0.46 | 7.79 |
| 30% | Mean-Std | 18.61 | 0.14 | 0.37 | 6.21 |
| | VaR95 | 18.86 | 0.15 | 0.32 | 6.36 |
| | CVaR95 | 24.58 | 0.12 | 0.31 | 5.65 |
| 40% | Mean-Std | 18.88 | 0.18 | 0.40 | 7.18 |
| | VaR95 | 20.92 | 0.17 | 0.35 | 7.13 |
| | CVaR95 | 23.38 | 0.15 | 0.35 | 6.64 |
| 50% | Mean-Std | 19.95 | 0.12 | 0.32 | 6.02 |
| | VaR95 | 19.62 | 0.14 | 0.40 | 7.13 |
| | CVaR95 | 24.95 | 0.23 | 0.35 | 8.96 |

## 6. Conclusions and Further Work

We have illustrated how D4PG can be used in conjunction with quantile regression to produce hedging strategies that are consistent with the objective of the hedger. Our approach allows the agent take transaction costs into account when developing a policy for managing gamma and vega. Our results illustrate that RL agent is able to find a good balance between (a) only hedging delta and (b) fully hedging gamma or vega as well as delta. Our robustness tests show that the hedging strategies we have developed are fairly robust to parameter estimates.

There are a number of ways in which our research can be extended. It would be interesting to explore how well the hedging strategies work for other stochastic volatility processes for the underlying asset and for processes where there are jumps in the price of the underlying asset. Up to now we have assumed that all the options traded with clients are "plain vanilla" European options. Further research could test how well our results carry over to exotic options such as barrier options. Also, we have used a fairly simple set up where a trade in a single at-the-money option is used for hedging each day. Alternative strategies where two or more options are available could be considered. More extensive robustness tests where the true process followed by the asset price is quite different from the assumed process could be carried out. The bid-ask spread, which determines the transaction costs could be assumed to be stochastic. Other tests could involve using convex or "fixed plus variable" transaction costs. The impact of the time horizon considered (30 days in our tests) could also be evaluated.

The economics of trading derivatives is an important concern to dealers. In the example we considered, we have shown that the profits on trades more than cover the costs of hedging. It would be interesting to carry out further tests to see how this result depends on the average number of client orders per day and the transaction costs.

# REFERENCES

Barth-Maron, G., M. W. Hoffman, D. Budden, W. Dabney, D. Horgan, D. TB, A. Muldal, N. Heess, and T. Lillicrap (2018), "Distributed distributional deterministic policy gradients," arXiv: 1804.08617.

Bellemare, M. G., W. Dabney, and R. Munos (2017), "A distributional perspective on reinforcement learning," arXiv: 1707.06887.

Black, F. and M. Scholes (1973), "The pricing of options and corporate liabilities," *Journal of Political Economy*, 81(May/June): 637-659.

Buehler H., L. Gonon, J. Teichmann and B. Wood (2019), "Deep hedging", *Quantitative Finance*, 19:8, 1271-1291. arXiv: 1802.03042

Cao, J., J. Chen, J. Hull, and Z. Poulos (2021), "Deep hedging of derivatives using reinforcement learning," *Journal of Financial Data Science*, Winter 3 (1): 10-27. arXiv: 2103.16409.

Dabney, W., M. Rowland, M. G. Bellemare, and R. Munos. (2018). Distributional Reinforcement Learning With Quantile Regression. *Proceedings of the AAAI Conference on Artificial Intelligence*, *32* (1). arXiv: 1710.10044.

Hagan, P., D. Kumar, A. S. Lesniewski, and D. E. Woodward (2002), "Managing smile risk," *Wilmott*, September: 84-108.

Halperin, I (2017), "QLBS: Q-Learner in the Black-Scholes(-Merton) worlds," arXiv 1712.04609.

Kolm, P. N and G. Ritter (2019), "Dynamic replication and hedging: a reinforcement learning approach," *Journal of Financial Data Science*, Winter 2 (1):159-171.

Merton, R. C. (1973), "Theory of rational option pricing," *Bell Journal of Economics and Management* Science, 4 (Spring):141-183.

Tamar, A, D. D. Castro, and S. Mannor (2016), "Learning the variance of the reward-to-go," *Journal of Machine Learning*, 17 (13): 1-36.